\documentstyle[aps,preprint,floats,psfig]{revtex}
\begin{document}
\newcommand {\be}{\begin{equation}}
\newcommand {\ee}{\end{equation}}
\newcommand {\bea}{\begin{eqnarray}}
\newcommand {\eea}{\end{eqnarray}}
\newcommand {\nn}{\nonumber}

\draft
%
%
%
%

\title{Quasi-Particle Bound States around Impurities in 
$\rm d_{x^2 - y^2}$-wave Superconductors
}

\author{Stephan Haas and Kazumi Maki}
\address{Department of Physics and Astronomy, University of Southern
California, Los Angeles, CA 90089-0484
}

\date{\today}
\maketitle

\begin{abstract}
Zn and Ni impurities in the hole-doped high-temperature superconductors
are known to have strong effects on thermodynamic and transport properties.
A recent scanning tunneling microscope study of Zn-doped
Bi2212 (Pan {\it et al.}, Nature {\bf 403}, 746 (2000)) has provided 
high-resolution images of the local density of states around 
non-magnetic impurities in
$\rm d_{x^2 - y^2}$-wave superconductors. These pictures contain detailed
information about the spinor wave functions $u({\bf r})$ and 
$v({\bf r})$ of bound states with energy $E_0 \sim \Delta/30$,
centered at
the Zn-sites. We show that this type of wave function follows from
the solutions of the Bogoliubov-de Gennes equations for 
$\rm d_{x^2 - y^2}$-wave superconductors.

\end{abstract}

{\it Introduction:}
Impurity-doping of anisotropic superconductors has turned out to 
be a valuable probe, setting them apart from conventional superconductors
with an s-wave order parameter. For example, doping of 
high-temperature superconductors, such as YBCO, LSCO, and BCCO,
with non-magnetic impurities has contributed greatly to establish the
underlying $\rm d_{x^2 - y^2}$-wave order parameter in these systems. 
\cite{hirschfeld,sun,haas}
In particular, an analysis of the thermodynamic and transport properties
in these compounds suggests that 
Zn-impurities can be modeled with a scattering potential in the 
unitary limit.\cite{sun,maki} On the other hand, 
little is known experimentally about the
local real-space structure of the impurity bound states around the 
Zn-sites, in spite of numerous theoretical studies on this question.
\cite{byers,salkola,tsuchiura} Very recently, Pan {\it et al.} have
provided detailed scanning tunneling microscope (STM) images of the
local tunneling conductivity, fixed at $\pm E_0$, where 
$E_0 \simeq \Delta/30$ is the binding energy of the impurity bound state 
around a Zn site in Bi2212. The corresponding wave function exhibits 
a fourfold symmetry, associated with the underlying $\rm CuO_2$ lattice.

In the context of conventional s-wave superconductors, this type of
bound state, in this case associated with a magnetic impurity,
has been predicted by Shiba.\cite{shiba} Recently, this idea has found 
experimental support by STM studies in the vicinity of Fe and Gd
impurity atoms on the surface of Nb superconductors.\cite{yazdani}

The objective of this paper is to present a simple solution of the
Bogoliubov-de Gennes equations for $\rm d_{x^2 - y^2}$-wave
superconductors\cite{degennes,morita} in the presence of a single impurity.
This corresponds to an experimental situation where the impurity
concentration is dilute enough, such that interactions between the 
impurities can be neglected. Within this approach,
we obtain a wave function which resembles very accurately the fourfold
symmetry which has recently been observed around Zn-impurities in
Bi2212\cite{pan},
where the energies of the associated bound states are very small
($|E_0|/\Delta \ll 1$). We believe that this image of the local
tunneling conductivity around the Zn-site, as seen by STM, provides 
strong evidence for $\rm d_{x^2 - y^2}$-wave symmetry. Therefore,
this method can also be used to test anisotropic order parameters 
in organic superconductors, and may serve as a straightforward tool
to explore the underlying symmetry in other new superconductors where
the nature of the order parameter is still under debate.

We also study the case of Ni impurities which may be modeled by a 
weaker scattering potential, close to the Born limit.\cite{puchkaryov}
In this case, we find that although $|v({\bf r})|^2$ is not much different
from the Zn-case,  $|u({\bf r})|^2$ is rotated by $\pi/4$ with respect 
to the strong-scattering limit corresponding to Zn.
Finally, we find that for impurities with an impurity potential
of intermediate strength, the lowest harmonic in $|u({\bf r})|^2$
is suppressed, and therefore a dominant eight-fold symmetry may be 
expected instead of the four-fold symmetry, observed in the weak and 
strong scattering limits.

{\it Bogoliubov-de Gennes equations and bound state wave functions:}
The Bogoliubov-de Gennes (BdG) equations for $\rm d_{x^2 - y^2}$-wave 
superconductors in the continuum limit are given by\cite{degennes,morita}
\bea
E u({\bf r}) &=& \left( -\frac{\nabla^2}{2 m} - \mu - V ({\bf r}) \right)
u({\bf r}) + \frac{1}{p_F^2} \Delta (\partial_x^2 - \partial_y^2 )
v({\bf r}),\\
E v({\bf r}) &=& - \left( -\frac{\nabla^2}{2 m} - \mu - V ({\bf r}) \right)
v({\bf r}) + \frac{1}{p_F^2} \Delta (\partial_x^2 - \partial_y^2 )
u({\bf r}),
\eea
where $\mu$ is the chemical potential, and $V ({\bf r}) > 0$ is the
impurity potential, centered at the site ${\bf r} = 0$. 

Let us first consider the case of Zn impurities where the energy of the
bound states is known to be very small\cite{pan}, $E_0 \simeq 0$. In the
following, we use a variational ansatz for the solutions of the BdG equations:
\bea
u({\bf r}) &=&  A \exp{(- \gamma r)} \left( J_0(p_F r) + \sqrt{2} \beta 
J_4(p_F r) \cos{(4 \phi)} \right),\\
 v({\bf r}) &=&  \sqrt{2} A \alpha \exp{(- \gamma r)} J_2(p_F r) \cos{(2 \phi)},
\eea
where $J_l(z)$ are Bessel functions of the first kind, and $p_F$ is the 
Fermi momentum. 
($A$ is the global normalization 
factor for the wave functions which can be neglected.)
Note that for Bi2212 it is believed that $p_F \simeq 0.7 \AA^-1$. This 
implies that $p_F a  \simeq 2$.\cite{chiao} The coefficients
$\alpha$, $\beta$, and $\gamma$ are determined variationally.  
Inserting Eqs. 3 and 4 into Eqs. 1 and 2, we find
\bea
E &=& K - V -\frac{1}{\sqrt{2}}\Delta \alpha, \nn \\
E \alpha &=& - K \alpha - \frac{1}{\sqrt{2}}\Delta(1+\frac{\beta}{\sqrt{2}}),\\
E \beta &=& K \beta - \frac{1}{2} \Delta \alpha \nn ,
\eea
where
\bea
K &\equiv &  \frac{\int_0^{\infty} dr r \left[
\left( \partial_r \exp{(-\gamma r)}
J_l(p_F r)\right)^2 + \left( l \exp{(-\gamma r)} J_l(p_F r)/r \right)^2
\right]}
{2m \int_0^{\infty} dr r \left( \exp{(-\gamma r)} J_l(p_F r)\right)^2} 
-\mu
\simeq \frac{\gamma^3}{m p_F},\\
V &\equiv & \frac{\int_0^{\infty} dr r \exp{(-2 \gamma r)} J_0^2(p_F r)
V({\bf r})}
{\int_0^{\infty} dr r \exp{(-2 \gamma r)} J_0^2(p_F r)} 
\simeq (2 \pi \gamma p_F) \int_0^{\infty} dr r \exp{(-2 \gamma r)}
J_0^2(p_F r) V ({\bf r}). 
\eea
In general, $K$ in Eq. 6 depends on which Bessel function $J_l(p_F r)$
is used. However, in the limit $\gamma/p_F \ll 1$, $K$ reduces to 
$\gamma^3/(m p_F)$ for all $J_l(p_F r)$ with $l \ll p_F/\gamma$. 
In the approximation for $V$,
only the dominant s-wave component ($l$ = 0) of the scattering potential 
$V({\bf r})$ has been considered. 
In the usual convention, which is used in this paper,
a positive sign of $V({\bf r})$ corresponds to the 
attraction of electrons by the impurity. In impurity-doped
Bi2212\cite{pan} the charge carriers are holes, 
and hence the opposite sign has to be chosen.

For the case of a Zn-impurity (strong scattering limit), we may assume
that $E \simeq 0$.\cite{pan} This gives $K \simeq V/2$,
$V \simeq 3\Delta/2 \approx \sqrt{2}\Delta$, 
$\alpha \simeq \sqrt{2}(V - K )/\Delta$,
and $\beta \simeq -\sqrt{2}$.
These choices yield the approximate parameter set 
$\alpha \simeq 1$, $\beta \simeq -1/\sqrt{2}$, and $\gamma \simeq
p_F (\sqrt{2} p_F \xi)^{-1/3}
\simeq 0.3 p_F$. 
The tunneling conductance is given by\cite{gygi}
\bea
\frac{d I}{d V} ({\bf r}) \propto
{\rm sech}^2\left( \frac{eV - E_0}{2T} \right) |u({\bf r})|^2
+ {\rm sech}^2\left( \frac{eV + E_0}{2T} \right) |v({\bf r})|^2.
\eea
Thus, at small temperatures, the local tunneling conductance around
the impurity site is dominated by $|u({\bf r})|^2$ for a fixed binding energy
$E_0$ and by $|v({\bf r})|^2$ for $-E_0$.
In Figs. 1 and 2 images of $|u({\bf r})|^2$ and
$|v({\bf r})|^2$ are shown. Both $|u({\bf r})|^2$ and $|v({\bf r})|^2$ have a 
four-fold symmetry, and extend in the directions of the Cu-O bonds. This
strongly resembles the images seen by the STM experiments.\cite{pan} 
Furthermore, weaker higher-harmonic
satellite peaks are observed in $|u({\bf r})|^2$,
along the ($\pm \pi/4, \pm \pi/4)$ directions, in accordance with 
the experiments\cite{pan}. Note that with our sign convention for
the scattering potential $V({\bf r})$, the roles of 
$|u({\bf r})|^2$ and $|v({\bf r})|^2$ are interchanged for the hole-doped
cuprates. Thus, in Bi2212 our $|u({\bf r})|^2$ corresponds to a bound
state at $-E_0$ and $|v({\bf r})|^2$ to a bound state at $E_0$.

Now let us turn to the case of Ni impurities. Ni is considered to
be a weak scatterer. Therefore we may assume that  $V({\bf r}) \rightarrow
0$, and consequently $\alpha \simeq 2 \beta (K - E)/\Delta$ and 
$\beta \simeq \sqrt{2}(1 - 4(E^2 - K^2)/\Delta^2)^-1$. 
So, if $E^2 > K^2 + \Delta^2/4$
the coefficient
$\beta$ turns out to be positive, which is most likely the case for Ni.
Indeed, a sufficiently large magnitude of $\alpha$ implies that 
$E^2 \gg K^2 \simeq 0$
and $E\simeq \sqrt{3}\Delta /2$. 
This gives $\alpha \simeq -\sqrt{3/2}$, $\beta \simeq 1/\sqrt{2}$,
and $\gamma \simeq
0.1 p_F$. The squares of the wave functions $u({\bf r})$ and $v({\bf r})$
are shown in 
Figs. 3 and 4. Comparing with the corresponding images for the
strong scattering limit,
we note that $|v({\bf r})|^2$ is qualitatively similar in both
cases, whereas $|u({\bf r})|^2$ appears to be rotated by $\pi/4$ with 
respect to the case of Zn impurities.  

For impurities with a scattering potential $V({\bf r})$ of intermediate 
strength, the coefficient $\beta$ in the wave function $u({\bf r})$ is found
to be greatly enhanced. Consequently, the higher-harmonic term in
$u({\bf r})$ (Eq. 3), containing
a $\cos{(4\phi)}$-modulation dominates the tunneling response at positive
binding energies $+E_0$. In this case, an eight-fold instead of a
four-fold symmetry should be
observed in $|u({\bf r})|^2$, as shown in Fig. 5. On the other hand,
the shape of
$|v({\bf r})|^2$ is not affected. The weaker satellite peaks in Fig. 1
can be viewed as strong-coupling precursors of this phenomenon. It would be
interesting to test this particular prediction of an eight-fold
symmetry in $|u({\bf r})|^2$ experimentally, using appropriate candidate 
impurity atoms with intermediate scattering strengths.  
 
{\it Conclusions:} In summary, we observe that (i) using the
Bogoliubov-de Gennes equations for single impurities in 
$\rm d_{x^2 - y^2}$-wave superconductors in the continuum limit 
is a very useful approach.
(ii) Both Zn and Ni impurities produce bound states
with a four-fold symmetry, localized around the impurity sites. 
(iii) Impurities with intermediate scattering strength may have 
a bound state with a dominant eight-fold symmetry. The simple analysis
of the BdG equations we have presented in this work provides a
semi-quantitative picture of the spatial structure of bound states
in anisotropic superconductors.

Recent experiments suggest that the electron-doped high-temperature 
superconductors NCCO and PCCO also have a d-wave order parameter.
\cite{tsuei,kokales,prozorov}
On the other hand, the effects of Zn and Ni doping in the electron-doped
systems appear to be opposite from the hole-doped case, i.e. Ni impurities   
lead to a stronger suppression
 of the superconducting ordering temperature than 
Zn doping.\cite{dastuto}
It is therefore likely that the roles of Zn and Ni are
interchanged in the electron-doped high-$\rm T_c$ cuprates. 
Furthermore, there are indications that the $\rm \kappa-(ET)_2$-salts
have a d-wave superconducting order parameter as well. 
\cite{carrington,pinteric} Therefore, similar studies of impurity 
bound states in these compounds would be of great interest. 

We thank Hyekyung Won for useful discussions.
S. H. acknowledges the Zumberge foundation for financial support.

\newpage

\begin{figure}[h]
\centerline{\psfig{figure=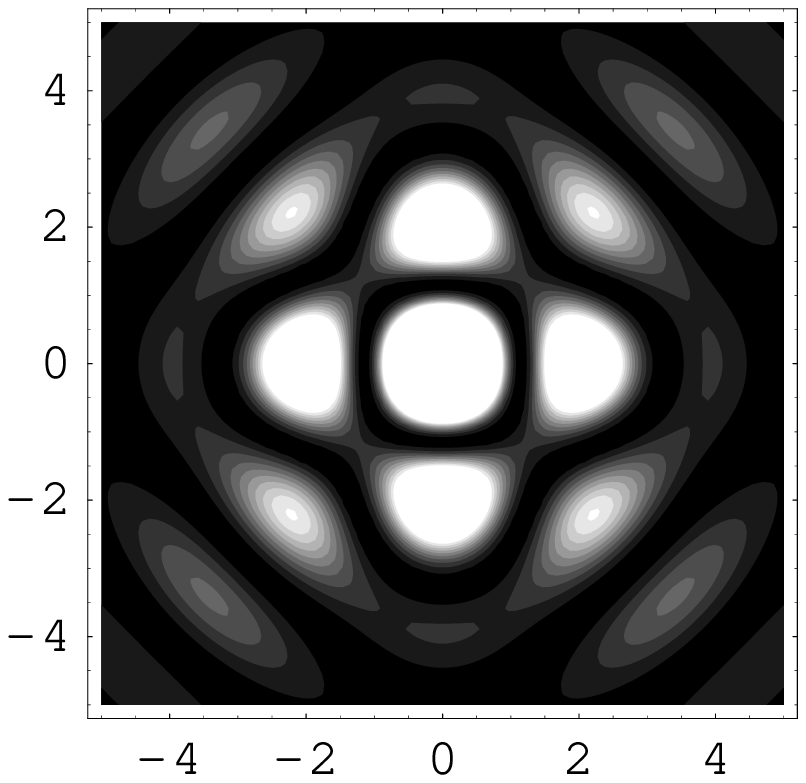,width=6cm,angle=0}}
\vspace{0.3cm}
\caption{
Spatial variation of the local tunneling conductance, centered at a
strong-scattering impurity, such as Zn, in a $\rm d_{x^2 - y^2}$-wave 
superconductor. In this figure, the dominant contribution 
$|u({\bf r})|^2$ at the 
positive bound state resonant frequency $+E_0$ is shown. 
}
\end{figure}

\begin{figure}[h]
\centerline{\psfig{figure=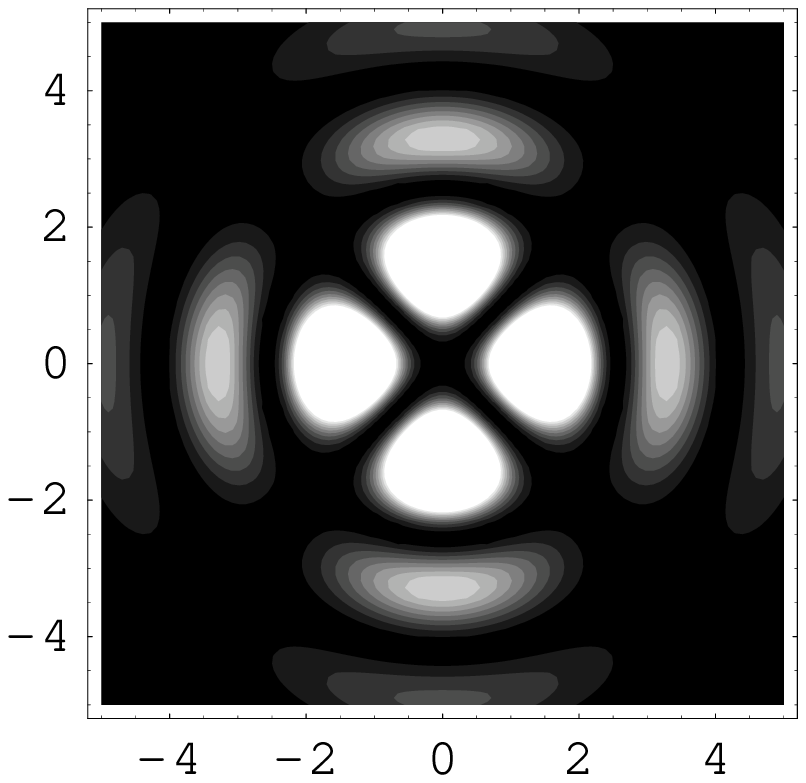,width=6cm,angle=0}}
\vspace{0.3cm}
\caption{
Spatial variation of the local tunneling conductance, localized around a
strong-scattering impurity, such as Zn, in a $\rm d_{x^2 - y^2}$-wave 
superconductor. Here the dominant contribution 
$|v({\bf r})|^2$ at the 
negative bound state resonant frequency $-E_0$ is shown. 
}
\end{figure}

\begin{figure}[h]
\centerline{\psfig{figure=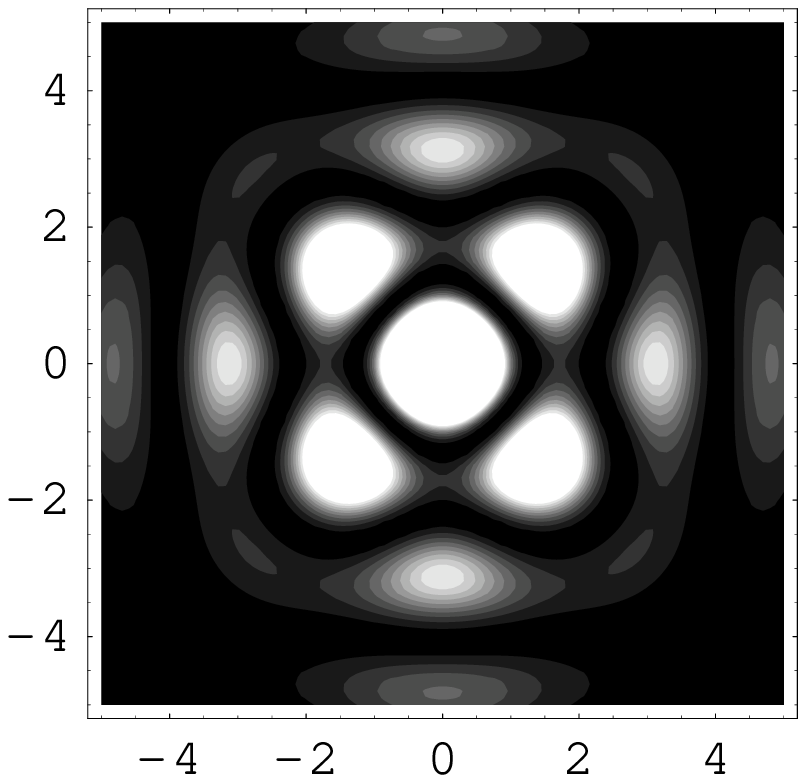,width=6cm,angle=0}}
\vspace{0.3cm}
\caption{
Spatial variation of the local tunneling conductance, centered at a
weak-scattering impurity, such as Ni, in a $\rm d_{x^2 - y^2}$-wave 
superconductor. In this figure, the dominant contribution 
$|u({\bf r})|^2$ at the 
positive bound state resonant frequency $+E_0$ is shown. 
}
\end{figure}

\begin{figure}[h]
\centerline{\psfig{figure=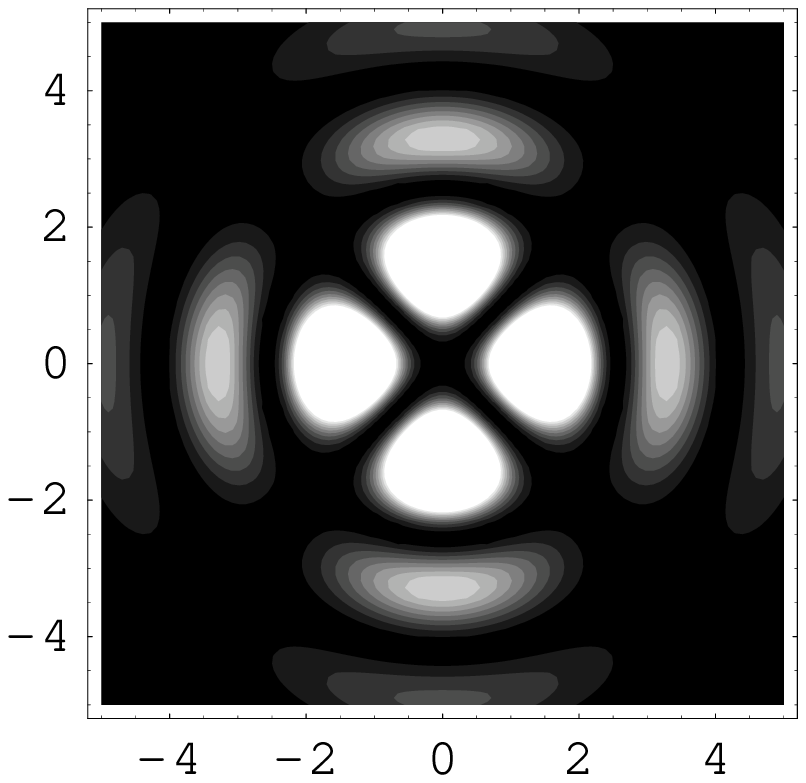,width=6cm,angle=0}}
\vspace{0.3cm}
\caption{
Spatial variation of the local tunneling conductance, localized around a
weak-scattering impurity, such as Ni, in a $\rm d_{x^2 - y^2}$-wave 
superconductor. Here the dominant contribution 
$|v({\bf r})|^2$ at the 
negative bound state resonant frequency $-E_0$ is shown. 
}
\end{figure}

\begin{figure}[h]
\centerline{\psfig{figure=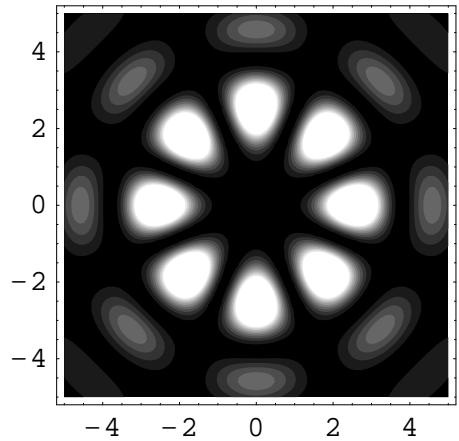,width=6cm,angle=0}}
\vspace{0.3cm}
\caption{
Spatial variation of the local tunneling conductance, centered at an
impurity of intermediate scattering strength. Here
the dominant contribution
$|u({\bf r})|^2$ at the 
positive bound state resonant frequency $+E_0$ is shown.
In contrast to the weak and strong scattering limits,
a dominant eight-fold symmetry is observed.
}
\end{figure}

\end{document}